\documentstyle[aps,tighten,preprint,epsfig,amssymb]{revtex}

\begin{document}

\draft

\preprint{nucl-th/0006057}

%%%%%%%%%%%%%%%%%%%%% Title %%%%%%%%%%%%%%%%%%%%%%

\title{
Nucleon resonances in $\bbox{\omega}$ photoproduction
}

%%%%%%%%%%%%%%%%%%%% Authors %%%%%%%%%%%%%%%%%%%%%

\author{
Yongseok Oh,$^a$%
\footnote{E-mail address: yoh@phya.yonsei.ac.kr,
Address after Aug. 1, 2000: Institute of Physics and Applied Physics,
Yonsei University, Seoul 120-749, Korea}
Alexander I. Titov,$^{b}$%
\footnote{E-mail address: atitov@thsun1.jinr.ru}
and T.-S. H. Lee\,$^c$%
\footnote{E-mail address: lee@anlphy.phy.anl.gov}
}

%%%%%%%%%%%%%%%%%%%% Addresses %%%%%%%%%%%%%%%%%%%%%

\address{
$^a$
Institute of Physics, Academia Sinica, Nankang,
Taipei 11529, Taiwan \\
$^b$
Bogoliubov Laboratory of Theoretical Physics, JINR,
Dubna 141980, Russia
\\
$^c$
Physics Division, Argonne National Laboratory, Argonne,
Illinois 60439}

\maketitle

%%%%%%%%%%%%%%%%%%%% Abstract %%%%%%%%%%%%%%%%%%%%%

\begin{abstract}
The role of the nucleon resonances ($N^*$) in $\omega$ photoproduction
is investigated by using the resonance parameters predicted by Capstick
and Roberts [Phys. Rev. D {\bf 46}, 2864 (1992); {\bf 49}, 4570 (1994)]. 
In contrast with the previous investigations based on the $\mbox{SU}(6)
\times \mbox{O}(3)$ limit of the constituent quark model, the employed
$N^* \to \gamma N$ and $N^* \to \omega N$ amplitudes include the 
configuration mixing effects due to the residual quark-quark interactions.
The contributions from the nucleon resonances are found to be significant
relative to the non-resonant amplitudes in changing the differential cross
sections at large scattering angles and various spin observables.
In particular, we suggest that a crucial test of our predictions can be
made by measuring the parity asymmetry and beam-target double asymmetry 
at forward scattering angles. 
The dominant contributions are found to be from $N\frac32^+ (1910)$,
a missing resonance, and $N\frac32^- (1960)$ which is identified as
the $D_{13}(2080)$ of the Particle Data Group.
\end{abstract}


\section{Introduction}

The constituent quark models predict a much richer nucleon excitation
spectrum than what have been observed in pion-nucleon scattering
\cite{IK77-80}.
This has been attributed to the possibility that a lot of the predicted
nucleon resonances ($N^*$) could couple weakly to the $\pi N$ channel.
Therefore it is necessary to search for the nucleon excitations in other
reactions to resolve the so-called ``missing resonance problem.''
Electromagnetic production of vector mesons ($\omega,\rho,\phi$)
is one of such reactions and is being investigated experimentally, e.g.,
at  Thomas Jefferson National Accelerator Facility, ELSA-SAPHIR of Bonn,
GRAAL of Grenoble, and LEPS of SPring-8.

The role of the nucleon excitations in vector meson photoproduction was
studied recently by Zhao {\em et al.\/} \cite{ZLB98,Zhao99,ZDGS99}
using an effective Lagrangian
method within the $\mbox{SU}(6) \times \mbox{O}(3)$ constituent quark model.
With the meson-quark coupling parameters adjusted to fit the existing data,
they found that the single polarization observables are sensitive to the
nucleon resonances.

We are motivated by the predictions by Capstick and Roberts
\cite{Caps92,CR94}.
They started with a constituent quark model which accounts for the
configuration mixing due to the residual quark-quark interactions
\cite{GI85-CI86}.
The predicted baryon wave functions are considerably different from those
of the $\mbox{SU}(6) \times \mbox{O}(3)$ model employed by Zhao
{\em et al.\/} in Refs. \cite{ZLB98,Zhao99,ZDGS99}.
The second feature of the predictions from Refs. \cite{Caps92,CR94} is
that the meson decays are calculated from the correlated wave functions
by using the ${}^3P_0$ model \cite{Yaou88}.
Thus it would be interesting to see how these predictions differ from
those of Refs. \cite{ZLB98,Zhao99,ZDGS99} and can be tested against the
data of vector meson photoproduction.

We will focus on $\omega$ photoproduction in this work, simply
because its non-resonant reaction mechanisms are much better understood.
It was fairly well established \cite{AB68,SS68,FRAA72,BCEK73,JLMS77}
already during the years around 1970 that this reaction is dominated by
diffractive processes at high energies and by one-pion exchange at
low energies.
The diffractive part can be described by the Pomeron exchange model.
The calculation of the one-pion exchange amplitude has been recently
revived by Friman and Soyeur \cite{FS96}.
It is therefore reasonable to follow the earlier theoretical analyses
\cite{FRAA72} and assume that the non-resonant amplitude of $\omega$
photoproduction can be calculated from these two well-established
mechanisms with some refinements.
The resulting model then can be a starting point for investigating
the $N^*$ effects.
This approach is similar to the previous investigation by Zhao
{\em et al.\/} \cite{ZLB98,Zhao99}.

In Sec.~II, we give explicit expressions for the non-resonant
amplitudes employed in our calculations.
The calculations of resonant amplitudes from Refs. \cite{Caps92,CR94}
are detailed in Sec.~III and the results are presented in Sec.~IV.
Section V is devoted to discussing possible future developments.

\section{Non-Resonant Amplitudes}

We assume that the non-resonant amplitude is due to the Pomeron exchange
[Fig.~\ref{fig:diag}(a)], pseudoscalar-meson exchange
[Fig.~\ref{fig:diag}(b)], and the direct and crossed nucleon terms
[Figs.~\ref{fig:diag}(c) and \ref{fig:diag}(d)]. 
The four-momenta of the incoming photon, outgoing $\omega$, initial
nucleon, and final nucleon are denoted as $k$, $q$, $p$, and $p'$
respectively, which defines $t = (p - p')^2 = (q-k)^2$,
$s \equiv W^2 = (p+k)^2$, and the $\omega$ production angle $\theta$
by $\cos\theta \equiv {\bf k} \cdot {\bf q} / |{\bf k}| |{\bf q}|$.

We choose the convention \cite{bd64} that the scattering amplitude $T$
is related to the $S$-matrix by
\begin{equation}
S^{}_{fi} = \delta^{}_{fi}
- i(2\pi)^4 \delta^4(k + p - q - p') T^{}_{fi}
\label{S:conv}
\end{equation}
with 
\begin{equation}
T^{}_{fi} = \frac{1}{(2\pi)^6} \frac{1}{\sqrt{2E^{}_\omega({\bf q})}}
\sqrt{\frac{M^{}_N}{E^{}_N({\bf p}')}} I^{}_{fi}
\frac{1}{\sqrt{2|{\bf k}|}} \sqrt{\frac{M^{}_N}{E^{}_N({\bf p})}},
\label{T:conv}
\end{equation}
where $E^{}_\alpha({\bf p}) = \sqrt{M^2_\alpha + {\bf p}^2}$ with
$M^{}_\alpha$ denoting the mass of the particle $\alpha$.
The invariant amplitude can be written as
\begin{eqnarray}
I^{}_{fi} = I^{bg}_{fi} + I^{N^*}_{fi},
\end{eqnarray}
where the nonresonant (background) amplitude is
\begin{eqnarray}
I^{bg}_{fi} = I^P_{fi} + I^{ps}_{fi} + I^{N}_{fi}
\end{eqnarray}
with $I^P_{fi}$, $I^{ps}_{fi}$, and $I^{N}_{fi}$ denoting the amplitudes
due to the Pomeron exchange, pseudoscalar-meson exchange, and direct and
crossed nucleon terms, respectively.
The nucleon excitation term $I^{N^*}_{fi}$ will be given in Sec.~III.

For the Pomeron exchange, which governs the total cross sections and
differential cross sections at low $|t|$ in the high energy region,
we follow the Donnachie-Landshoff model \cite{DL84-92},
which gives \cite{LM95,PL97}
\begin{equation}
I_{fi}^P = i M^{}_0 (s,t) \, \bar{u}^{}_{m_f^{}} (p')
\varepsilon^*_{\mu}(\omega)
\left\{ k \!\!\!/ g^{\mu\nu} - k^\mu \gamma^\nu \right\}
\varepsilon_{\nu}(\gamma) u^{}_{m_i^{}}(p),
\end{equation}
where $\varepsilon^{}_{\mu} (\omega)$ and $\varepsilon^{}_{\nu} (\gamma)$
are the polarization vectors of the $\omega$ meson and photon,
respectively, and $u^{}_m(p)$ is the Dirac spinor of the nucleon with
momentum $p$ and spin projection $m$.
The Pomeron exchange is described by the following Regge parameterization
\begin{equation}
M^{}_0 (s,t)= C^{}_V \, F^{}_1(t) \, F^{}_V (t)\,
\left(\frac{s}{s_0} \right)^{\alpha_P^{} (t)-1}
\exp\left\{ - \frac{i\pi}{2} [ \alpha_P^{}(t) - 1] \right\},
\label{Pom:M0}
\end{equation}
where $F_1^{}(t)$ is the isoscalar electromagnetic form factor of the
nucleon and $F_V^{}(t)$ is the form factor for the
vector-meson--photon--Pomeron coupling.
We also follow Ref. \cite{DL84-92} to write
\begin{eqnarray}
F_1^{} (t) &=& \frac{ 4 M_N^2-2.8t }{ (4M_N^2-t) (1-t/t_0)^2 },
\nonumber \\
F_V^{} (t) &=& \frac{1}{1 - t/M_V^2}\,
\frac{2\mu_0^2}{2 \mu_0^2 + M_V^2 - t},
\end{eqnarray}
where $t_0 = 0.7$ GeV$^2$.
The Pomeron trajectory is known to be $\alpha_P^{} (t) = 1.08 + 0.25\,t$.
(See also Ref. \cite{CKK97})
The strength factor $C_V^{}$ reads $C_V^{} = 12 \sqrt{4\pi\alpha_{\rm em}}
\beta_0^2/f_V^{}$ with the vector meson decay constant $f_V$ ($=17.05$
for the $\omega$ meson) and $\alpha^{}_{\rm em} = e^2/4\pi$.
By fitting all of the total cross section data for $\omega$, $\rho$, and
$\phi$ photoproduction at high energies, the remaining parameters of the
model are determined:
$\mu_0^2 = 1.1$ GeV$^2$, $\beta_0^{} = 2.05$ GeV$^{-1}$, and
$s_0 = 4$ GeV$^2$.

The pseudoscalar-meson exchange amplitude can be calculated
from the following effective Lagrangians,
\begin{eqnarray}
{\cal L}_{\omega \gamma \varphi}^{} &=&
\frac{e g_{\omega\gamma \varphi}}{M_V}
\epsilon^{\mu\nu\alpha\beta}
\partial_\mu \omega_\nu \partial_\alpha A_\beta\, \varphi,\qquad
\nonumber\\
{\cal L}_{\varphi NN}^{} &=&
-i g_{\pi NN} \bar N \gamma_5\tau_3 N \pi^0
-i g_{\eta NN} \bar N \gamma_5 N \eta,
\end{eqnarray}
where $\varphi = (\pi^0,\eta)$ and $A_\beta$ is the photon field.
The resulting invariant amplitude is
\begin{eqnarray}
I^{ps}_{fi} = - \sum_{\varphi=\pi,\eta}
\frac{iF_{\varphi NN}(t) F_{\omega\gamma\varphi}(t)}{t-M_\varphi^2}
\frac{e g^{}_{\omega\gamma\varphi} g^{}_{\varphi NN}}{M_V} \,
\bar{u}_{m^{}_f}(p') \gamma_5 u_{m^{}_i}(p) \,
\varepsilon^{\mu\nu\alpha\beta} q_{\mu}^{} k^{}_\alpha
\varepsilon_\nu^* (\omega) \varepsilon^{}_\beta (\gamma).
\label{T:ps}
\end{eqnarray}
In the above, we have followed Ref. \cite{FS96} to include the following
form factors to dress the $\varphi NN$ and $\omega\gamma\varphi$
vertices,
\begin{equation}
F_{\varphi NN}^{} (t) = \frac{\Lambda_\varphi^2 - M^2_\varphi}
{\Lambda_\varphi^2 -t},  \qquad
F_{\omega\gamma\varphi}^{} (t) =
\frac{\Lambda_{\omega\gamma\varphi}^2-M_\varphi^2}
{\Lambda_{\omega\gamma\varphi}^2-t} .
\label{PS:FF}
\end{equation} 
We use $g_{\pi NN}^2/4\pi = 14$ for the $\pi NN$ coupling constant.
The $\eta NN$ coupling constant is not well determined \cite{TBK94}.
Here we use $g^2_{\eta NN}/4\pi = 0.99$ which is obtained from
making use of the SU(3) symmetry relation \cite{deS63} together with
a recent value of $F/D = 0.575$ \cite{CR93}.
The coupling constants $g_{\omega\gamma\varphi}$ can be estimated
through the decay widths of $\omega \to \gamma\pi$ and
$\omega \to \gamma \eta$ \cite{PDG98} which lead to
$g_{\omega\gamma\pi} = 1.823$ and $g_{\omega\gamma\eta} = 0.416$.
The cutoff parameters $\Lambda_\varphi$ and
$\Lambda_{\omega\gamma\varphi}$ in Eq. (\ref{PS:FF}) will be specified
in Sec.~IV.

We evaluate the direct and crossed nucleon amplitudes shown in
Fig.~\ref{fig:diag}(c) and \ref{fig:diag}(d) from the following interaction
Lagrangians,
\begin{eqnarray}
{\cal L}_{\gamma NN}^{} & = &
- e \bar{N} \left( \gamma_\mu \frac{1+\tau_3}{2} {A}^\mu
- \frac{\kappa_N^{}}{2M_N^{}} \sigma^{\mu\nu} \partial_\nu A_\mu \right) N,
\nonumber\\
{\cal L}_{\omega NN}^{} & = &
- g_{\omega NN}^{} \bar{N} \left( \gamma_\mu {\omega}^\mu
- \frac{\kappa_\omega}{2M_N^{}} \sigma^{\mu\nu}
\partial_\nu \omega_\mu \right) N,
\end{eqnarray}
with the anomalous magnetic moment of the nucleon $\kappa_{p(n)} =
1.79$ $(-1.91)$.
There are some uncertainties in choosing the $\omega NN$ coupling constants.
In this work, we consider $g_{\omega NN}^{} = 7.0 \sim 11.0$ and
$\kappa_\omega \approx 0$, which are determined in a study of $\pi N$
scattering and pion photoproduction \cite{SL96}.
The resulting invariant amplitude reads
\begin{equation}
I^N_{fi} = \bar{u}_{m^{}_f} (p') \varepsilon^{\mu *} (\omega)
M_{\mu\nu} \varepsilon^\nu (\gamma) u_{m^{}_i}(p),
\label{T:N}
\end{equation}
where
\begin{eqnarray}
M_{\mu\nu} = -e g_{\omega NN}^{}
\left[ \Gamma_\mu^\omega(q)
\frac{\not\! p \ + \not\! k \ + M_N}{s-M_N^2}
\Gamma_\nu^\gamma(k) F_N^{} (s) +
\Gamma_\nu^\gamma(k)
\frac{\not\! p \ - \not\! q \ + M_N}{u-M_N^2}
\Gamma_\mu^\omega(q) F_N^{} (u)
\right]
\label{N-term}
\end{eqnarray}
with
\begin{equation}
\Gamma_\mu^\omega (q) = \gamma_\mu - i \frac{\kappa_\omega}{2M_N}
\sigma_{\mu\alpha} q^\alpha, \qquad
\Gamma_\nu^\gamma (k) = \gamma_\nu + i \frac{\kappa_p}{2M_N}
\sigma_{\nu\beta} k^\beta,
\label{Gamma-N}
\end{equation}
and $s= (p+k)^2$, $u = (p-q)^2$.
Here we have followed Ref. \cite{HBMF98a} to include a form factor    
\begin{eqnarray}
F_N (r) = \frac{\Lambda_N^4}{\Lambda_N^4  - (r - M_N^2)^2} 
\label{N:FF}
\end{eqnarray}
with $r = s$ or $t$.
The cutoff parameter $\Lambda_N$ will be specified later in Sec.~IV.

The amplitude (\ref{N-term}) is not gauge-invariant because
of the form factors $F_N (s)$ and $F_N (u)$.
(Note that the terminology ``gauge invariance'' here only means
the ``current conservation'' conditions $M_{\mu\nu} k^\nu = q^\mu
M_{\mu\nu} = 0$ as considered in most investigations.)
To restore the gauge invariance, we follow Ref. \cite{TOYM98} and
modify the amplitude $M_{\mu\nu}$ by using the projection operator
$P_{\mu\nu} = g_{\mu\nu} - k_\mu q_\nu / k \cdot q$;
\begin{equation}
M_{\mu\nu} \to P_{\mu\mu'} M^{\mu'\nu'} P_{\nu'\nu},
\label{trans1}
\end{equation}
which leads to the following modifications in evaluating the amplitude
(\ref{N-term}):
\begin{eqnarray}
\Gamma_\mu^\omega(q) &\to& \Gamma_\mu^\omega(q)
- \frac{1}{k\cdot q} k_\mu q \cdot \Gamma^\omega(q),
\nonumber \\
\Gamma_\nu^\gamma(k) &\to& \Gamma_\nu^\gamma(k)
- \frac{1}{k\cdot q} q_\nu k \cdot \Gamma^\gamma(k).
\label{trans2}
\end{eqnarray}
The above prescription is certainly very phenomenological, while
it is similar to other accepted approaches in literature.
Perhaps a more rigorous approach can be developed by extending the 
work for pseudoscalar meson production \cite{HBMF98a} to the
present case of vector meson production.
This is however beyond the scope of this investigation.
For our present exploratory purposes, the prescription defined by Eqs.
(\ref{trans1}) and (\ref{trans2}) is sufficient.

\section{Resonant Amplitude}

In order to estimate the nucleon resonance contributions we make use
of the quark model predictions on the resonance photo-excitation
$\gamma N \to N^*$ and the resonance decay $N^* \to \omega N$ reported
in Refs. \cite{Caps92,CR94} using a relativised quark model. 
The resonant amplitude is illustrated in Fig.~\ref{fig:diag}(e). 
The crossed $N^*$ amplitude, similar to Fig.~\ref{fig:diag}(d), cannot
be calculated from the informations available in Refs. \cite{Caps92,CR94}
and are not considered in this work.
Here we follow Ref. \cite{SL96} and write the resonant amplitude 
in the center of mass frame as [in the convention
defined by Eqs. (\ref{S:conv}) and (\ref{T:conv})]
\begin{eqnarray}
I^{N^*}_{m_f,m_\omega,m_i,\lambda_\gamma}({\bf q},{\bf k})
= \sum_{J,M_J^{}}
\frac{{\cal M}_{N^*\to N'\omega}({\bf q};m_f^{},m_\omega^{};J,M_J^{})
{\cal M}_{\gamma N \to N^*}({\bf k};m_i^{},\lambda_\gamma^{};J,M_J^{})}
{\sqrt{s} - M_R^J + \frac{i}{2}\Gamma^J(s)},
\label{T:N*}
\end{eqnarray}
where $M^J_R$ is the mass of a $N^*$ with spin quantum numbers
$(J, M_J)$, and $m_i$ , $m_f$, $\lambda_\gamma$, and $m_\omega$ are the
spin projections of the initial nucleon, final nucleon, incoming photon,
and outgoing $\omega$ meson, respectively.
Here we neglect the effect due to the nonresonant mechanisms on the
$N^*$ decay amplitudes and the shift of the resonance position.
Then the resonance mass $M_R^J$ and the $N^*$ decay amplitudes
${\cal M}_{N^* \to \gamma N, \omega N}$ can be identified with the quark
model predictions of Refs. \cite{Caps92,CR94}, as discussed in Refs.
\cite{SL96,YSAL00}. 
We however do not have information about the total decay width 
$\Gamma^J(s)$ for most of the $N^*$'s considered here.
For simplicity, we assume that its energy-dependence is similar to
the width of the $N^* \rightarrow \pi N$ decay within the oscillator
constituent quark model.
Following Ref. \cite{YSAL00} and neglecting the real part of the mass
shift, we then have
\begin{equation}
\Gamma^J (s) = \Gamma^J_0
\frac{\rho(k_\pi)}{\rho(k_{0\pi})}
\left( \frac{k_\pi}{k_{0\pi}} \right)^{2L_\pi}
\exp\left[2 ({\bf k}_{0\pi}^2 - {\bf k}_\pi^2)/\Lambda^2 \right],
\label{N*:decay}
\end{equation}
where $L_\pi$ is the orbital angular momentum of the 
considered $\pi N$ state and
\begin{equation}
\rho(k) = \frac{k E_N E_\pi}{E_N + E_\pi}.
\end{equation}
In the above equations, $k_\pi(\equiv |{\bf k}_\pi|)$ is the pion momentum
at energy $\sqrt{s}$ while $k_{0\pi}$ is evaluated at $\sqrt{s} = M^J_R$.
Our choice of the total average width $\Gamma^J_0$ and cutoff parameter
$\Lambda$ for Eq. (\ref{N*:decay}) will be specified in Sec.~IV.

By setting the photon momentum in the $z$-direction, the
$N^* \to \gamma N$ amplitudes in Eq. (\ref{T:N*}) can be calculated
from the helicity amplitudes $A_{\lambda}$ listed in
Ref. \cite{Caps92} from
\begin{equation}
{\cal M}_{\gamma N \to N^*}({\bf k};m_i^{},\lambda_\gamma^{};J,M_J^{})
= \sqrt{2k} \, A_{M_J^{}}^{}
\delta_{M_J^{}, \lambda_\gamma^{} + m_i^{}}
f({\bf k},{\bf k}^{}_0),
\label{photoN*}
\end{equation}
where ${\bf k}^{}_0$ is the photon momentum at the resonance position,
i.e.  at $\sqrt{s}=M^J_R$, and the factor $f({\bf k},{\bf k}^{}_0)$ was
introduced to evaluate the amplitude in the region where the photon
momentum is off the resonant momentum ${\bf k}^{}_0$.
To be consistent with the form factor of the total decay width
(\ref{N*:decay}), we set $f({\bf k},{\bf k}^{}_0) = \exp[({\bf k}_0^2
- {\bf k}^2)/\Lambda^2]$.

The $N^* \to \omega N$ amplitude takes the following form \cite{caps00}
\begin{eqnarray}
{\cal M}_{N^*\to N'\omega}({\bf q};m_f^{},m_\omega^{};J,M_J^{})
&=& 2\pi \sqrt{\frac{2M_R^J}{M_N^{}|{\bf q}_0^{}|}}
\sum_{L, m_L^{} S, m_S^{}}
\langle L\, m_L^{}\, S\, m_S^{} | J\, m_J^{} \rangle
\nonumber \\ && \times
\langle 1\, m_\omega^{}\, {\textstyle\frac12}\, m_f^{} | S\, m_S^{} \rangle
Y_{L m_L^{}}(\hat{q}) \, G(L,S)
\left( |{\bf q}|/|{\bf q}^{}_0| \right)^L f({\bf q},{\bf q}^{}_0),
\label{omegaN*}
\end{eqnarray}
where $G(L,S)$'s are listed in Refs. \cite{CR94,caps00}, and ${\bf q}_0$
is the $\omega$ meson momentum at $\sqrt{s} = M_R^J$.
Here we also include the extrapolation factor $f({\bf q},{\bf q}^{}_0) =
\exp[({\bf q}_0^2-{\bf q}^2)/\Lambda^2]$ like that for the
$N^* \to \gamma N$ vertex in Eq. (\ref{photoN*}).

In this study, we consider 12 positive parity and 10 negative parity
nucleon resonances up to spin-$9/2$.
The resonance masses $M^J_R$, the transition amplitudes $A_{M_J^{}}^{}$,
and $G(L,S)$ needed to evaluate Eqs. (\ref{T:N*})-(\ref{omegaN*}) are
taken from Refs. \cite{Caps92,CR94} and are listed in
Tables~\ref{tab:Nstar1} and \ref{tab:Nstar2}.
Three of them were seen in the $\pi N$ channel with four-star rating,
five of them with two-star rating, and one of them with one-star rating.
These $N^*$'s are indicated in the last column (PDG) of
Tables~\ref{tab:Nstar1} and \ref{tab:Nstar2}.
Clearly the majority of the predicted $N^*$'s are ``missing'' so far.
Here we should also mention that we are not able to account for the
resonances with its predicted masses less than the $\omega N$ threshold,
since their decay vertex functions with an off-shell momentum are
not available in Refs. \cite{Caps92,CR94}.

\section{Results and Discussions}

The model defined in Secs.~II and III involves some free parameters which
must be specified.
The Pomeron exchange model parameters, as given explicitly in Sec.~II,
are taken from a global fit to the total cross sections of $\omega$,
$\rho$, and $\phi$ photoproduction at high energies and will not be
adjusted in this study.
The Pomeron exchange becomes weak at low energies, as shown by the
dot-dashed curves in Fig.~\ref{fig:dsdt}.
We therefore will determine the parameters of the other amplitudes
mainly by considering the data at low energies.

The pseudoscalar-meson exchange amplitude in Eq. (\ref{T:ps}) depends
on the cutoff parameters $\Lambda_{\pi(\eta)}$ and
$\Lambda_{\omega\gamma\pi(\eta)}$ of Eq. (\ref{PS:FF}).
The $\eta$ exchange is very weak for any choice of its cutoff parameters.
For definiteness, we choose $\Lambda_\eta = 1.0$ GeV and 
$\Lambda_{\omega\gamma\eta} = 0.9$ GeV as determined in a 
study \cite{TLTS99} of $\phi$ photoproduction.
At low energies, the $\pi$ exchange completely dominates the cross
sections at forward angles.
Its cutoff parameters $\Lambda_\pi$ and $\Lambda_{\omega\gamma\pi}$ thus
can be fixed by fitting the forward cross section data.
Our best fit is obtained by setting $\Lambda_\pi = 0.6$ GeV and
$\Lambda_{\omega\gamma\pi} = 0.7$ GeV.  
These values are slightly different from those of Ref. \cite{FS96}.
The resulting contributions from the pseudoscalar-meson exchange are
the dashed curves in Fig.~\ref{fig:dsdt}.

The resonant amplitude defined by Eqs. (\ref{T:N*})-(\ref{omegaN*})
depends on the oscillator parameter $\Lambda$ and the averaged total
width $\Gamma^J_0$.
We find that our results are rather insensitive to the cutoff
$\Lambda$ in the range $\Lambda = 0.5 \sim 1.0$ GeV.
We take the value $\Lambda= 1$ GeV which is the value of the standard
harmonic oscillator constituent quark model \cite{IK77-80}.
For the averaged total width $\Gamma^J_0$, we are guided by the widths
listed by the Particle Data Group \cite{PDG98}.
For the $N^*$'s which have been observed and listed in the last column
(PDG) of Tables~\ref{tab:Nstar1} and \ref{tab:Nstar2}, their widths are
all very large in the range of about $200 \sim 400$ MeV \cite{PDG98,MS92}.
The other $N^*$'s considered in our calculations are expected to have
similar large widths.
We therefore choose the average of the values listed by PDG and set 
$\Gamma^J_0 = 300$ MeV for all $N^*$'s included in our calculation.
The resulting $N^*$ contributions are the dotted curves in
Fig.~\ref{fig:dsdt}.

With the pseudoscalar-meson exchange and resonant amplitudes fixed by
the above procedure, the parameters for the direct and crossed 
nucleon amplitude [Eqs. (\ref{T:N})-(\ref{N:FF})] are then adjusted to
fit the data.
Here we consider $g_{\omega NN} = 7 \sim 11$ and $\kappa_\omega = 0$ as
determined in a study of $\pi N$ scattering and $\gamma N \to \pi N$
reaction \cite{SL96}.
This range of $\omega NN$ coupling constant is very close to
$g_{\omega NN} = 10.35$ determined \cite{RSY99} recently from fitting
the nucleon-nucleon scattering data.
Thus the only free parameter in the fit is the cutoff parameter
$\Lambda_N$ of the form factor in Eq. (\ref{N:FF}).
It turns out that the contributions from the direct and crossed nucleon
terms are backward peaked, and $\Lambda_N$ can be fairly well determined
by total cross sections at backward scattering angles.
Our best fits are obtained from setting $\Lambda_N = 0.5$ GeV with
$g_{\omega NN} = 10.35$ and $\kappa_\omega = 0$.
The corresponding contributions from the direct and crossed nucleon
terms are the dot-dot-dashed curves in Fig.~\ref{fig:dsdt}.

Our full calculations including all amplitudes illustrated in
Fig.~\ref{fig:diag} are the solid curves in Fig.~\ref{fig:dsdt}.
The data can be described to a very large extent in the
considered energy region $ E_\gamma \leq 5 $ GeV.
It is clear that the contributions due to the $N^*$ excitations (dotted
curves) and the direct and crossed nucleon terms (dot-dot-dashed curves)
help bring the agreement with the data at large angles.
The forward angle cross sections are mainly due to the interplay between
the pseudoscalar-meson exchange (dashed curves) and the Pomeron exchange
(dot-dashed curves). 
The main problem here is in reproducing the data at $E_\gamma = 1.23$ GeV.
This perhaps indicates that the off-shell contributions from $N^*$'s
below $\omega N$ threshold are important at very low energies.
These sub-threshold $N^*$'s cannot be calculated from the informations
available so far within the model of Refs. \cite{Caps92,CR94} and are
neglected in our calculations. 
The investigation of this possibility is however beyond the scope of this
work.
The quality of our fit is comparable to that of Zhao {\em et al.\/}
\cite{ZLB98}.

\begin{figure}[h]
\centering
\epsfig{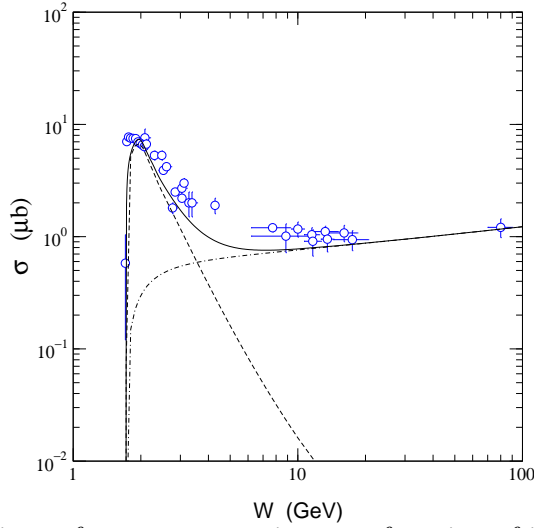}
\caption{Total cross sections of $\gamma p\to p\omega$ reaction
as a function of invariant mass $W$.
The solid curve is from the full calculation and the dotted curve is
from the calculation without including Pomeron exchange.
The Pomeron exchange contribution is given by the dot-dashed line.
Data are taken from Refs. \protect\cite{BCEK73,Klein96-98,Durham}.}
\label{fig:totcs}
\end{figure}

It is important to note here that various cutoff parameters determined
above also fix the high energy behavior of our predictions.
Thus the accuracy of our model must be tested by examining whether we are
able to describe the total cross sections from threshold to very high
energies.
Our prediction (solid curves) are compared with the available data in
Fig.~\ref{fig:totcs}.
We see that our model indeed can reproduce the data very well except in
the region close to $W = 5$ GeV where the Pomeron exchange (dot-dashed curve)
and the sum of the other amplitudes (dashed curve) are comparable.
It is interesting to note here that if we increase the Pomeron exchange
strength $C_V$ of Eq. (\ref{Pom:M0}) by about 10$\%$, the total cross
section data can be much better described without too much changes in
describing the low energy data.
However, the Pomeron exchange parameters are constrained by a global fit
to all of data for $\rho$, $\omega$, and $\phi$ photoproduction, and
therefore such a change is not desirable.
Instead, we must explore other mechanisms, such as the absorption effects
due to the intermediate $N\rho$ state as discussed in Ref. \cite{SS68}.
Since the $N^*$ excitations considered here are in the region $W \leq 2.5$
GeV, we need not to resolve the problem in this transition region near
$W = 5$ GeV.

\begin{figure}[h]
\centering
\epsfig{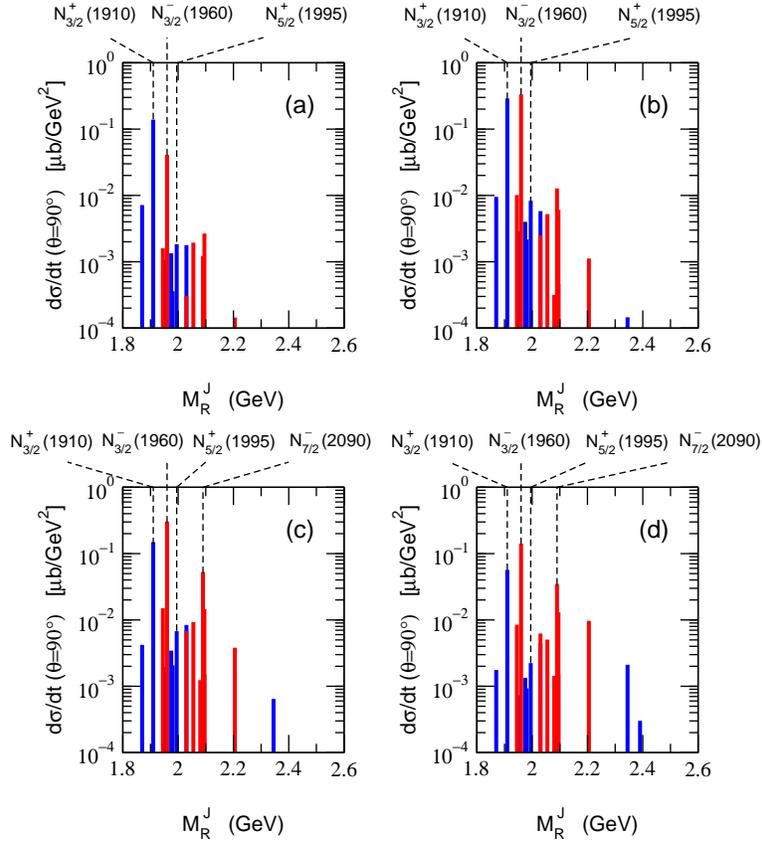}
\bigskip
\caption{Contribution from each $N^*$ listed in Tables \ref{tab:Nstar1}
and \ref{tab:Nstar2} to the differential cross sections at
$\theta = 90^\circ$ and $E_\gamma = $
(a) $1.23$, (b) $1.45$, (c) $1.68$, and (d) $1.92$ GeV,
which corresponds to $W = $ (a) $1.79$, (b) $1.90$, (c) $2.01$,
and (d) $2.11$ GeV, respectively.}
\label{fig:N*}
\end{figure}

To have a better understanding of the resonance contributions, we
compare in Fig.~\ref{fig:N*} the contributions from the considered
$N^*$'s to the differential cross sections at $\theta = 90^\circ$. 
Here the $N^*$ states listed in Tables~\ref{tab:Nstar1} and
\ref{tab:Nstar2} can be identified by its mass $M^J_R$. 
As also indicated in Fig.~\ref{fig:N*}, the contributions from
$N\frac32^+ (1910)$ and $N\frac32^- (1960)$ are the largest at all
energies.
{}From Tables I and II, we see that the $N\frac32^+ (1910)$ is a missing
resonance, while $N\frac32^- (1960)$ is identified by Capstick
\cite{Caps92a} as a two star resonance $D_{13}(2080)$ of PDG. 
In the study of Zhao et al. \cite{ZLB98,Zhao99}, they found that
$F_{15}(2000)$ dominates.
This resonance is identified with $N{\frac52^+}(1995)$ in Table I and is
found to be not so strong in our calculation, as also indicated in
Fig.~\ref{fig:N*}.  
This significant difference between the two calculations is not surprising
since the employed quark models are rather different.
In particular, our predictions include the configuration mixing effects
due to residual quark-quark interactions.

In Fig.~\ref{fig:N*} we also see that the relative importance between
different resonances depend on the photon energy.
As expected from the resonance part expression (\ref{T:N*}), the higher
mass resonances become more important as the photon energy increases
from $1.23$ GeV ($W=1.79$ GeV) to $1.92$ GeV ($W=2.11$ GeV).
For example, we also indicate in Fig.~\ref{fig:N*} that the contribution
from $N\frac72^- (2090)$, identified in Table II with $G_{17}(2190)$ of
PDG, becomes comparable to that of $N\frac32^+ (1910)$ at $W=2.11$ GeV.

\begin{figure}[h]
\centering
\epsfig{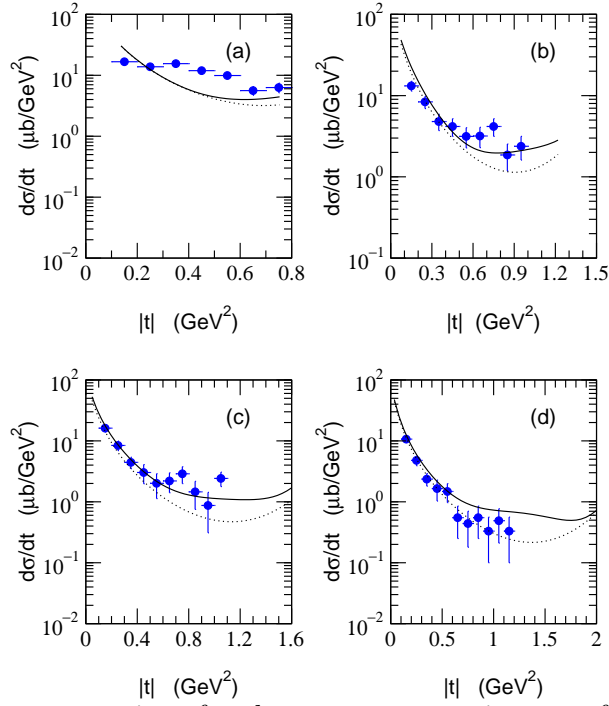}
\caption{
Differential cross sections for the $\gamma p\to p\omega$ reaction
as a function of $|t|$ at different energies: $E_\gamma =$ (a) $ 1.23$,
(b) $1.45$, (c) $1.68$, and (d) $1.92$ GeV. 
The solid and dotted curves are calculated respectively with and without
including $N^*$ effects.
Data are taken from Ref. \protect\cite{Klein96-98}.}
\label{fig:dsN*}
\end{figure}

The total resonance effects are shown in Fig.~\ref{fig:dsN*}.
The solid curves are from our full calculations, while the dotted curves
are from the calculations without including $N^*$ excitations.
The results shown in Fig.~\ref{fig:dsN*} indicate that it is rather
difficult to test our predictions by considering only the angular
distributions, since the $N^*$'s influence is mainly in the large
scattering angle region where accurate measurements are perhaps still
difficult.
On the other hand, the forward cross sections seem to be dominated 
by the well-understood pseudoscalar-meson exchange (less-understood
$\eta$ exchange is negligibly small here) and Pomeron exchange.
Therefore, one can use this well-controlled background to examine the
$N^*$ contributions by exploiting the interference effects in the spin
observables.

\begin{figure}[h]
\centering
\epsfig{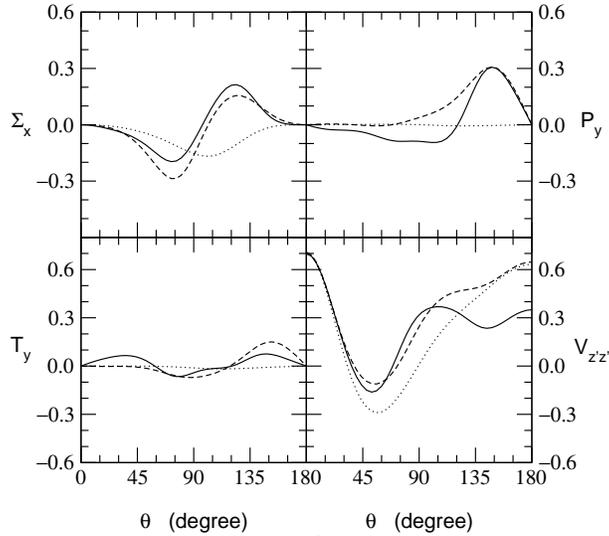}
\caption{
Single asymmetries at $E_\gamma = 1.7$ GeV.
The dotted curves are calculated without including $N^*$ effects, the
dashed curves include contributions of $N\frac32^+(1910)$ and
$N\frac32^-(1960)$ only, and the solid curves are calculated with all
$N^*$'s listed in Tables \ref{tab:Nstar1} and \ref{tab:Nstar2}.}
\label{fig:single}
\end{figure}

We first examine the spin observables discussed in Refs.
\cite{ZLB98,Zhao99}.
Our predictions for photon asymmetry ($\Sigma_x$), target asymmetry
($T_y$), recoil nucleon asymmetry ($P_y$), and tensor polarization
($V_{z'z'}$) are shown in Fig.~\ref{fig:single}.
These single polarization observables are calculated according to the
definitions given, e.g., in Refs. \cite{TOYM98,PST96}.
We see that the $N^*$ excitations can change the predictions from the
dotted curves to solid curves.
The dashed curves are obtained when only the $N\frac32^+(1910)$ and
$N\frac32^-(1960)$ are included in calculating the resonant part of the
amplitude.
Our predictions are significantly different from those of Refs.
\cite{ZLB98,Zhao99}.
As mentioned above, this is perhaps mainly due to the differences between
the employed quark models.
Nevertheless, we confirm their conclusion that the single polarization
observables are sensitive to the $N^*$ excitations but mostly at large
scattering angles.
The vector asymmetry $V_y$ has also been investigated, but is not presented
here since it is almost impossible to access experimentally \cite{KCT98}.

\begin{figure}[h]
\centering
\epsfig{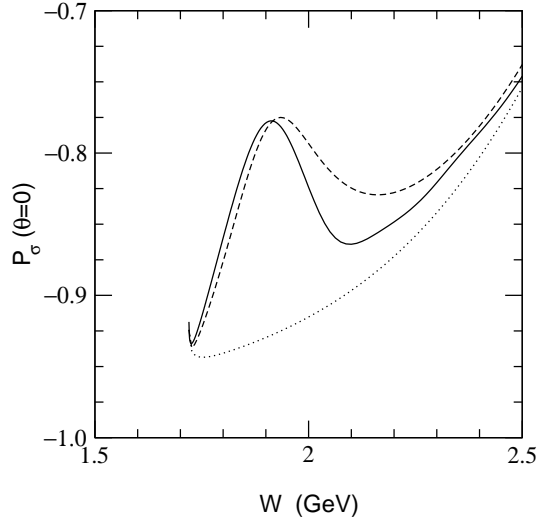}
\caption{
Parity asymmetry $P_\sigma$ at $\theta=0$ as a function of $W$.
Notations are the same as in Fig. \ref{fig:single}.}
\label{fig:Psigma}
\end{figure}

To further facilitate the experimental tests of our predictions, we have
also investigated other spin observables.
We have identified two polarization observables which are sensitive
to the $N^*$ contributions at forward angles, where precise measurements
might be more favorable because the cross sections are peaked at
$\theta = 0$ (see Fig.~\ref{fig:dsdt}).
The first one is the parity asymmetry defined as \cite{SSW}
\begin{equation}
P_\sigma = 
\frac{d\sigma^N - d\sigma^U}{d\sigma^N + d\sigma^U}
= 2\rho^1_{1-1}-\rho^1_{00},
\end{equation} 
where $\sigma^N$ and $\sigma^U$ are the cross sections due to the
natural and unnatural parity exchanges respectively, and
$\rho^{i}_{\lambda,\lambda^\prime}$ are the vector-meson spin
density matrices.
For the dominant one-pion exchange amplitude, which has unnatural
parity exchange only, one expects $P_\sigma = -1$. 
Thus any deviation from this value will be only due to $N^*$ excitation
and Pomeron exchange, since the contribution from the direct and crossed
nucleon terms is two or three orders in magnitude smaller at
$\theta = 0$ (see Fig.~\ref{fig:dsdt}).
Our predictions for $P_\sigma$ are shown in Fig.~\ref{fig:Psigma}.
We show the results from calculations with (solid curve) and without
(dotted curve) including $N^*$ contributions.
The difference between them is striking and can be unambiguously tested
experimentally.
In the considered low energy region, most of the $N^*$ excitations
involve both the natural and unnatural parity exchanges.
The rapid energy-dependence of the solid curve thus reflects the change
of relative importance between different $N^*$'s as energy increases, as
seen in Fig.~\ref{fig:N*}.
At $W \ge 2.5$ GeV, the Pomeron exchange starts to dominate and shift
the prediction to the $P_\sigma = +1$ limit of natural parity exchange.
Here we also find that the $N\frac32^+ (1910)$ and $N\frac32^- (1960)$
are dominant.
By keeping only these two resonances in calculating the resonant part of
the amplitude, we obtain dashed curve which is not too different from
the full calculation (solid curve).

\begin{figure}[h]
\centering
\epsfig{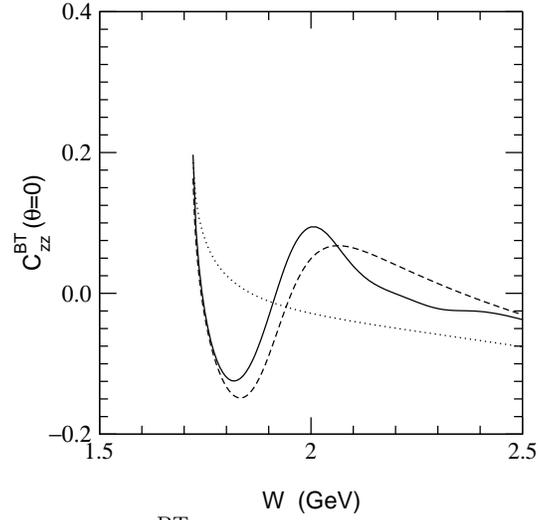}
\caption{
Beam-target asymmetry $C^{\rm BT}_{zz}$ at $\theta=0$ as a function of $W$.
Notations are the same as in Fig. \ref{fig:single}.}
\label{fig:BT}
\end{figure}

We also find that the beam-target double asymmetry at forward angles is
sensitive to the $N^*$ excitations.
It is defined as \cite{TOYM98}
\begin{equation}
C^{BT}_{zz} = 
\frac{d\sigma(\uparrow\downarrow) - d\sigma(\uparrow\uparrow)}
{d\sigma(\uparrow\downarrow) + d\sigma(\uparrow\uparrow)},
\end{equation}
where the arrows represent the helicities of the incoming photon and
the target protons.
In Fig.~\ref{fig:BT}, we present our predictions on $C^{BT}_{zz}$
at $\theta=0$ as a function of invariant mass $W$.
The striking difference between the solid curve and dotted curve is due to
the $N^*$ excitations.
At $W \geq $ 2.5 GeV this asymmetry vanishes since all amplitudes except
the helicity-conserving natural parity Pomeron exchange are suppressed
at high energies.
The role of $N^*$ here is similar to what is discussed above for 
$P_\sigma$.
Again, the $N\frac32^+ (1910)$ and $N\frac32^- (1960)$ give the dominant
contributions.
This is evident from comparing the solid curve and the dashed curve which
is obtained when only these two resonances are kept in calculating the
resonant part of the amplitude.
It is very interesting to note that the $N\frac32^- (1960)$ is also
found to be important in kaon photoproduction \cite{MB99}, although its 
identification with the $D_{13}(2080)$ is still controversial.
Our predicions show that $\omega$ photoproduction can be useful
in resolving this issue.

\section{Future Developments}

In this work, we have investigated the role of nucleon resonances in
$\omega$ photoproduction.
The resonance parameters are taken from the predictions of
Refs. \cite{Caps92,CR94}.
It is found that the resonant contributions can influence significantly
the differential cross sections at large angles.
We have presented predictions showing the $N^*$ effects on several spin
observables.
In particular, we have shown that our predictions can be crucially
tested by measuring the parity asymmetry and beam-target double
asymmetry at $\theta = 0$. 
The dominant contributions are found to be from $N\frac32^+ (1910)$,
a missing resonance, and $N\frac32^- (1960)$ which was indentified
\cite{Caps92} as the two star resonance $D_{13}(2080)$ of PDG.
Experimental test of our predictions will be a useful step toward
resolving the so-called ``missing resonance problem'' or distinguishing
different quark model predictions.

\begin{figure}
\centering
\epsfig{file=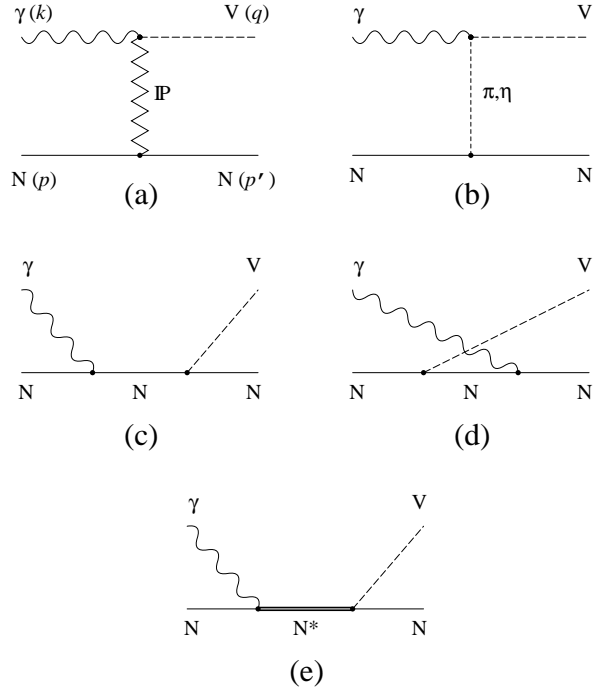, width=8cm}
\caption{
Diagrammatic representation of $\omega$ photoproduction mechanisms:
(a) Pomeron exchange, (b) ($\pi,\eta$) exchange, (c) direct nucleon
term, (d) crossed nucleon term, and (e) $s$-channel nucleon excitations.}
\label{fig:diag}
\end{figure}

\begin{figure}
\centering
\epsfig{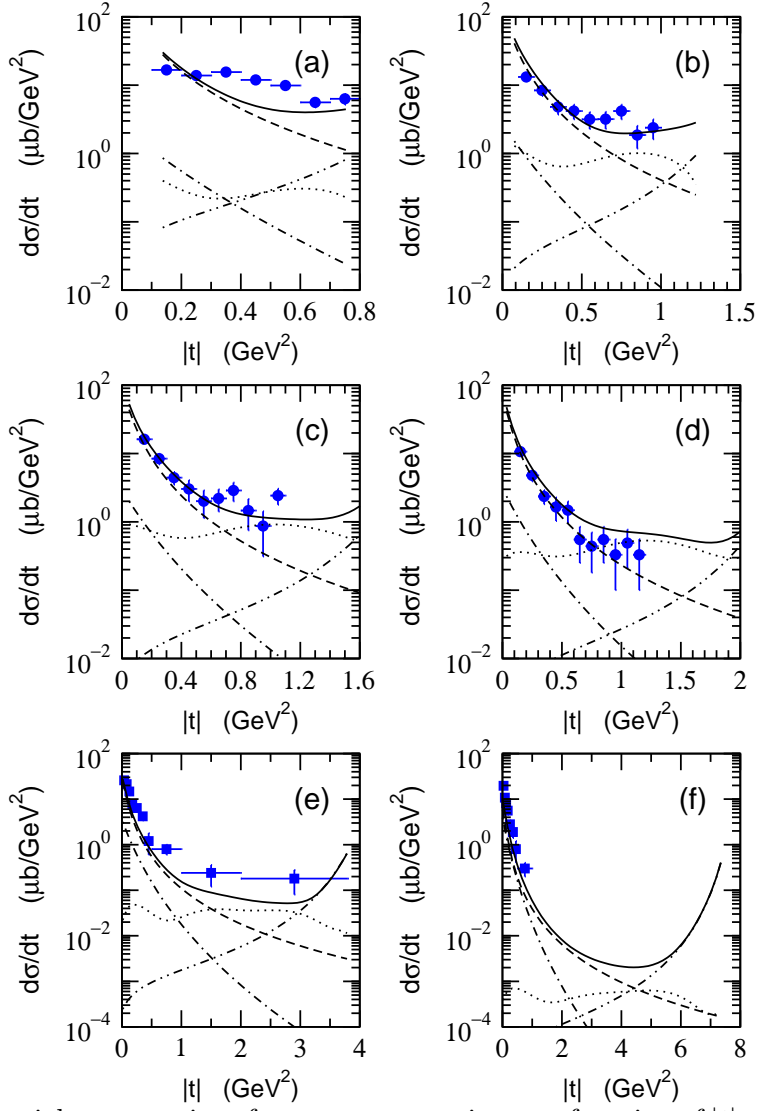}
\caption{
Differential cross sections for $\gamma p\to p\omega$ reaction
as a function of $|t|$ at $E_\gamma =$ (a) $1.23$, (b) $1.45$, (c)
$1.68$, (d) $1.92$, (e) $2.8$, and (f) $4.7$ GeV.
The results are from pseudoscalar-meson exchange (dashed),
Pomeron exchange (dot-dashed), direct and crossed nucleon terms
(dot-dot-dashed), $N^*$ excitation (dotted), and the full amplitude (solid).
Data are taken from Ref. \protect\cite{Klein96-98} (filled circles)
and Ref. \protect\cite{BCEK73} (filled squares).}
\label{fig:dsdt}
\end{figure}

To end, we should emphasize that the present investigation is a very
first step from the point of view of a dynamical treatment of the problem,
as has been done for the $\pi N$ scattering and pion photoproduction
\cite{SL96,YSAL00}.
The main uncertain part is the lack of a complete calculation of the
$N^* \to \gamma N, \omega N$ transition form factors given in
Eqs. (\ref{photoN*}) and (\ref{omegaN*}).
The use of the extrapolation factor $f({\bf k},{\bf k}_0) =
\exp[({\bf k}^2_0 - {\bf k}^2)/\Lambda^2]$ must be justified by extending
the calculations of Refs. \cite{Caps92,CR94} to evaluate these form
factors for any off-shell momentum.
It is also needed to generate the form factors for the $N^*$'s which are
below $\omega N$ threshold and are neglected in this investigation.
The effects due to the sub-threshold $N^*$'s could be important in
explaining the data very close to $\omega N$ threshold [e.g.,
Fig.~\ref{fig:dsdt}(a) at $E_\gamma= 1.23$ GeV].
An another necessary step is to develop an approach to calculate the
crossed $N^*$ amplitude [similar to the crossed nucleon amplitude
Fig.~\ref{fig:diag}(d)] using the same relativised constituent quark
model employed in Refs. \cite{Caps92,CR94}.
Finally, the effects due to the initial and final state interactions must
be also investigated, which may be pursued by extending the approach of
Ref. \cite{SS68}.

%%%%%%%%%%%%%%%%%  Acknowledgment  %%%%%%%%%%%%%%%%%%%%%%%

\acknowledgements

This work was supported in part by National Science Council of Republic
of China, Russian Foundation for Basic Research under Grant No. 96-15-96426, 
and U.S. DOE Nuclear Physics Division Contract No. W-31-109-ENG-38.

%%%%%%%%%%%%%%%%%  References  %%%%%%%%%%%%%%%%%%%%%%%

%%%%%%%%%%%%%%%%%  Tables  %%%%%%%%%%%%%%%%%%%%%%%

\begin{table}
\centering
\begin{tabular}{ccccccccc} 
%%%%%%%%%%%%%%%%% spin 1/2
$N^*$& $M^J_R$ & $A_{1/2}$ & $A_{3/2}$ &
$G(1,1/2)$ &$G(1,3/2)$ &            & $\sqrt{\Gamma^{\rm
tot}_{N\omega}}$ & PDG \cite{PDG98} \\ \hline
$N\frac12^+$ & $1880$ & $0$   &  ---  & $-4.3$ & $-1.6$ & ---    &
$4.6$ & \\ 
$N\frac12^+$ & $1975$ & $-12$ &  ---  & $-3.1$ & $-0.8$ & ---    &
$3.1$ & \\ \hline
%%%%%%%%%%%%%%%%% spin 3/2
  & & & & $G(1,1/2)$ &$G(1,3/2)$ & $G(3,3/2)$ & & \\ \hline
$N\frac32^+$ & $1870$ & $-2$  & $-15$ & $0.0$  & $+4.4$ & $+0.6$ &
$4.5$ & $P_{13} (1900)^{\star\star}$ \\ 
$N\frac32^+$ & $1910$ & $-21$ & $-27$ & $-5.8$ & $+5.7$ & $-0.5$ &
$8.2$ & \\
$N\frac32^+$ & $1950$ & $-5$  & $2$   & $-5.4$ & $-3.2$ & $+0.7$ &
$6.3$ & \\
$N\frac32^+$ & $2030$ & $-9$  & $15$  & $-1.6$ & $-2.9$ & $+0.7$ &
$3.3$ & \\
\hline
%%%%%%%%%%%%%%%%% spin 5/2
  & & & & $G(3,1/2)$ &$G(1,3/2)$ & $G(3,3/2)$ & & \\ \hline
$N\frac52^+$ & $1980$ & $-11$ & $-6$  & $+2.1$ & $-1.7$ & $-1.1$ &
$2.9$ & \\
$N\frac52^+$ & $1995$ & $-18$ & $1$   & $-0.3$ & $+3.1$ &
$-1.6$ & $3.5$ & $F_{15} (2000)^{\star\star}$ \\
\hline
%%%%%%%%%%%%%%%%% spin 7/2
  & & & & $G(3,1/2)$ &$G(3,3/2)$ & $G(5,3/2)$ & & \\ \hline
$N\frac72^+$ & $1980$ & $-1$  & $-2$  & $-0.8$ & $+1.4$ & $0.0$
& $1.6$ & $F_{17} (1990)^{\star\star}$ \\
$N\frac72^+$ & $2390$ & $-14$ & $-11$& $-0.8$ & $+2.1$ & $+2.0$ & $3.0$
& \\
$N\frac72^+$ & $2410$ & $+1$ & $-1$& $-0.7$ & $+1.3$ & $0.0$ & $1.5$ & \\
\hline
%%%%%%%%%%%%%%%%% spin 9/2
  & & & & $G(5,1/2)$ &$G(3,3/2)$ & $G(5,3/2)$ & & \\ \hline
$N\frac92^+$ & $2345$ & $-29$ & $+13$& $-0.3$ & $-2.9$ & $-0.6$ &
$2.9$ & $H_{19} (2220)^{\star\star\star\star}$ \\
\end{tabular}
\caption{
Parameters for positive parity nucleon resonances from Refs.
\protect\cite{Caps92,CR94}.
The helicity amplitude $A_\lambda$ is given in unit of
$10^{-3}$ GeV$^{-1/2}$.
$G(L,S)$ and 
$\sqrt{\Gamma^{\rm tot}_{N\omega}}$ are in unit of MeV$^{1/2}$.
The resonance mass $M^J_R$ is in unit of MeV.}
\label{tab:Nstar1}
\end{table}

\begin{table}
\centering
\begin{tabular}{ccccccccc} 
%%%%%%%%%%%%%%%%% spin 1/2
$N^*$& $M^J_R$ & $A_{1/2}$ & $A_{3/2}$ &
$G(0,1/2)$ &$G(2,3/2)$ & & $\sqrt{\Gamma^{\rm tot}_{N\omega}}$ & PDG
\cite{PDG98} \\ \hline
$N\frac12^-$ & $1945$ & $+12$ & --- & $-0.9$ & $-5.6$ &  & $5.7$ &
$S_{11} (2090)^{\star}$ \\
$N\frac12^-$ & $2030$ & $+20$ & --- & $-0.1$ & $-2.8$ &  & $2.8$ & \\
\hline
%%%%%%%%%%%%%%%%% spin 3/2
 & & & & $G(2,1/2)$ &$G(0,3/2)$ & $G(2,3/2)$ & & \\ \hline
$N\frac32^-$ & $1960$ & $+36$ & $-43$ & $-4.3$ & $-0.2$ & $-4.6$ &
$6.3$ & $D_{13} (2080)^{\star\star}$ \\
$N\frac32^-$ & $2055$ & $+16$ & $0$ & $+2.0$ & $-1.3$ & $-2.7$ & $3.6$ & \\
$N\frac32^-$ & $2095$ & $-9$ & $-14$ & $-3.2$ & $+1.9$ & $+3.8$ & $5.3$
& \\
\hline
%%%%%%%%%%%%%%%%% spin 5/2
 & & & & $G(2,1/2)$ &$G(2,3/2)$ & $G(4,3/2)$ & & \\ \hline
$N\frac52^-$ & $2080$ & $-3$ & $-14$ & $-2.2$ & $-0.3$ & $+2.0$ & $2.9$
& \\
$N\frac52^-$ & $2095$ & $-2$ & $-6$ & $-3.1$ & $+3.3$ & $+0.8$ &
$4.6$ & $D_{15} (2200)^{\star\star}$ \\
\hline
%%%%%%%%%%%%%%%%% spin 7/2
 & & & & $G(4,1/2)$ &$G(2,3/2)$ & $G(4,3/2)$ & & \\ \hline
$N\frac72^-$ & $2090$ & $-34$ & $+28$ & $-1.5$ & $-3.7$ & $-1.7$ &
$4.4$ & $G_{17} (2190)^{\star\star\star\star}$ \\
$N\frac72^-$ & $2205$ & $-16$ & $+4$ & $-0.2$ & $-5.1$ & $+0.3$ & $5.1$
& \\
\hline
%%%%%%%%%%%%%%%%% spin 9/2
 & & & & $G(4,1/2)$ &$G(4,3/2)$ & $G(6,3/2)$ & & \\ \hline
$N\frac92^-$ & $2215$ & $0$ & $+1$ & $-1.0$ & $+1.7$ & $0.0$ &
$2.0$ & $G_{19} (2250)^{\star\star\star\star}$ \\
\end{tabular}
\caption{
Parameters for negative parity nucleon resonances from Refs.
\protect\cite{Caps92,CR94}.
The units are the same as in Table \ref{tab:Nstar1}.}
\label{tab:Nstar2}
\end{table}

%%%%%%%%%%%%%%%%%  Figures  %%%%%%%%%%%%%%%%%%%%%%%

\end{document}